\documentclass[aps,prx,notitlepage,superscriptaddress,longbibliography,twocolumn,nofootinbib,floatfix]{revtex4-2}

\usepackage[utf8]{inputenc}
\usepackage{graphicx}
\usepackage{hyperref}
\usepackage{xcolor,graphicx}

\usepackage{tikz,tikz-3dplot}
\def\checkmark{\tikz\fill[scale=0.4](0,.35) -- (.25,0) -- (1,.7) -- (.25,.15) -- cycle;} 

\usepackage{lipsum}

\usepackage{amsmath,amsthm,amssymb,mathtools}
\usepackage{multirow}
\usepackage{braket}
\usepackage{bbm,bm}
\usepackage[bb=boondox]{mathalfa}

\DeclareFontFamily{U}{matha}{\hyphenchar\font45}
\DeclareFontShape{U}{matha}{m}{n}{
      <5> <6> <7> <8> <9> <10> gen * matha
      <10.95> matha10 <12> <14.4> <17.28> <20.74> <24.88> matha12
      }{}
\DeclareSymbolFont{matha}{U}{matha}{m}{n}
\DeclareMathSymbol{\muparrow}{3}{matha}{"D2}
\DeclareMathSymbol{\mdownarrow}{3}{matha}{"D3}
\DeclareMathSymbol{\mupdownarrow}{3}{matha}{"D9}

\DeclareFontFamily{U}{mathb}{\hyphenchar\font45}
\DeclareFontShape{U}{mathb}{m}{n}{
      <5> <6> <7> <8> <9> <10> gen * mathb
      <10.95> mathb10 <12> <14.4> <17.28> <20.74> <24.88> mathb12
      }{}
\DeclareSymbolFont{mathb}{U}{mathb}{m}{n}
\DeclareMathSymbol{\mdownuparrows}{3}{mathb}{"D7}

\usepackage{makecell}

\newcommand{\ie}{{\it i.e.},\ }
\newcommand{\eg}{{\it e.g.},\ }

\begin{document}
\title{Hilbert space fragmentation and interaction-induced localization in the extended Fermi-Hubbard model}

\author{Philipp Frey}
\email{pfrey@student.unimelb.edu.au}
\affiliation{School of Physics, The University of Melbourne, Parkville, VIC 3010, Australia}

\author{Lucas Hackl}
\email{lucas.hackl@unimelb.edu.au}
\affiliation{School of Mathematics and Statistics, The University of Melbourne, Parkville, VIC 3010, Australia}
\affiliation{School of Physics, The University of Melbourne, Parkville, VIC 3010, Australia}

\author{Stephan Rachel}
\email{stephan.rachel@unimelb.edu.au}
\affiliation{School of Physics, The University of Melbourne, Parkville, VIC 3010, Australia}

	\begin{abstract}
		We study Hilbert space fragmentation in the extended Fermi-Hubbard model with nearest and next-nearest-neighbor interactions. Using a generalized spin/mover picture and saddle point methods, we derive lower bounds for the scaling of the number of frozen states and for the size of the largest block preserved under the dynamics. We find fragmentation for strong nearest- and next-nearest-neighbor repulsions as well as for the combined case. Our results suggest that the involvement of next-nearest-neighbor repulsions leads to an increased tendency for localization. We then model the dynamics for larger systems using Markov simulations to test these findings and unveil in which interaction regimes the dynamics becomes spatially localized. In particular, we show that for strong nearest- and next-nearest-neighbor interactions random initial states will localize provided that the density of initial movers is sufficiently low.
	\end{abstract}
	
	\maketitle
	
	Dynamical thermalization and ergodicity breaking in closed many-body quantum systems have been subjects of renewed interest in recent years. Advances in controlled experimental and quantum simulation techniques have allowed us to observe quantum dynamics with unprecedented resolution and even to engineer entirely novel many-body Hamiltonians\,\cite{PhysRevB.76.052203,Schreiber842,Smith2016,Zhang2017,PhysRevLett.121.023601,Smith_2019,kyprianidis-21s1192,Mi2022,doi:10.1126/sciadv.abm7652}. This experimental progress is accompanied by new theoretical insights into the emergence of thermodynamic properties from quantum dynamics. In addition, possible exceptions to what seems to be the generic mechanism of thermalization have been under investigation in recent years\,\cite{doi:10.1146/annurev-conmatphys-031214-014726,PhysRevB.82.174411,Wilczek_2012,PhysRevB.88.014206,RevModPhys.91.021001,PhysRevLett.118.030401,PhysRevB.94.085112,srednicki-94pre888,khemani-16prl250401}.
	A thermal Hamiltonian is characterized by eigenstates that look thermal with respect to local observables and have additional properties that ensure that arbitrary initial states thermalize, as long as they are not unphysical. These properties are summarized in what is called the \textit{eigenstate thermalization hypothesis} (ETH)\,\cite{srednicki-94pre888,deutsch91pra2046,rigol-08n854}. One of the proposed exceptions to ETH is many-body localization, which can be understood as an emergent integrability caused by local quenched disorder\,\cite{PhysRevB.75.155111, PhysRevB.82.174411, RevModPhys.91.021001, ALET2018498, PhysRevB.91.081103, PhysRevLett.113.107204, doi:10.1146/annurev-conmatphys-031214-014701,PhysRevB.98.174202}. A different mechanism is quantum many-body scars\,\cite{bernien-17n579,moudgalya-18prb235155,moudgalya-18prb235156,turner-18np745,iadecola-19prl036403,lin-19prl173401,serbyn-21np675,Moudgalya_2022}. These are a set of eigenstates of measure zero in the thermodynamic limit that violate ETH. A third mechanism of ergodicity breaking, that of \textit{Hilbert space fragmentation}, describes the phenomenon that local constraints can separate the Hilbert space into exponentially many subspaces that are spanned by product states and are dynamically disconnected from each other\,\cite{Moudgalya_2022, PhysRevB.101.214205, PhysRevX.12.011050, Moudgalya_2021, PhysRevX.10.011047, PhysRevLett.127.150601, PhysRevB.101.174204, PhysRevX.9.021003, PhysRevB.103.L220304}. The Hamiltonian assumes a block-diagonal structure with respect to a product basis, and the number of blocks is exponential in system size $L$. This results in a small number of accessible states and can prevent a typical initial product state from thermalizing. One may distinguish strong and weak fragmentation by comparing the dimension $\mathcal{D}_\mathrm{max}$ of the largest block to the dimension $\mathcal{D}$ of the entire Hilbert space. The former is characterized by  $\mathcal{D}_\mathrm{max}/\mathcal{D} \to 0$; the latter is characterized by $\mathcal{D}_\mathrm{max}/\mathcal{D} \to 1$ as $L \to \infty$. One striking feature of fragmentation is the abundance of so-called \textit{frozen states}, i.e., local product states that are also eigenstates of the Hamiltonian and hence evolve trivially.
	The most prominent class of systems that have been shown to be fragmented are dipole moment-conserving models in which only local multi-particle hopping terms are present\,\cite{PhysRevX.9.021003,PhysRevX.10.011047,PhysRevB.101.174204,Moudgalya_2022}. The nature of these constraints can lead to spatial localization of the dynamics in the sense that the occupancy of certain sites remains fixed over time\,\cite{PhysRevB.101.214205}. This is accompanied by a transition from weak to strong fragmentation. Another route to fragmentation relies on interaction-induced local constraints\,\cite{Dias_2000, de2019dynamics}. It was shown that the integrable spinless Fermi-Hubbard chain features a fragmented Hilbert space in the infinite interaction strength limit\,\cite{de2019dynamics}.
	
	In this Letter, we extend the spinless Hubbard model by adding a next-nearest-neighbor term and consider various limits and their effect on Hilbert space fragmentation and localization. Making use of a convenient mapping of product states onto symbol strings, we derive effective hopping rules for the novel cases. They lead us to construct frozen states as well as large Hilbert space blocks.
	Classical Markov simulations allow us to explore spatial localization in said limits beyond system sizes that can be diagonalized exactly\,\cite{PhysRevB.101.214205}. The effective rules are particularly simple for only nearest-neighbor constraints and combined nearest and next-nearest neighbor constraints. While the former case is clearly delocalized, our findings suggest that the additional next-nearest neighbor interactions induce localization.
	
	\textbf{Extended Fermi-Hubbard model.} We consider the $t$-$V_1$-$V_2$ spinless fermionic chain with periodic boundary conditions (PBCs) imposed,
	\begin{equation}
		\begin{aligned}
			\hat{H}\!=\!-t\!\sum_x (\hat{c}^\dagger_{x+1}\hat{c}_x\!+\!\mathrm{H.c.})\!+\! V_1\!\sum_{x}\hat{n}_x\hat{n}_{x+1}\! +\! V_2\!\sum_{x}\hat{n}_x\hat{n}_{x+2}.\label{eq:modelH}
		\end{aligned}
	\end{equation}
	$\hat{c}_x$ creates an electron on site $x$, and $t$ is the hopping amplitude. $V_1$ and $V_2$ denote nearest- and next-nearest-neighbor repulsion, respectively. We note that the model is integrable (nonintegrable) for $V_2=0$ ($V_2\not= 0$). Hamiltonian \eqref{eq:modelH} has been studied as a genuine fermionic model but also in its closely related hard-core boson form\,\cite{cazalilla2011one,leblond2021universality,bianchi2021volume}, as an anisotropic
	spin-$\frac{1}{2}$ XXZ chain\,\cite{orbach1958linear,walker1959antiferromagnetic} and the six-vertex model of statistical mechanics\,\cite{lieb1967residual}. The important limit $V_1\to \infty$ and $V_2=0$ was explicitly worked out for fermions using a spin/mover picture in\,\cite{Dias_2000}.
	Both limits ($V_1 \to \infty$, $V_2 = 0$) and ($V_2 \to \infty$, $V_2 = 0$) lead to a locally constrained hopping term:
	
	\begin{equation}
		\hat{H}^{(1,2)}_\infty = - \sum_x \hat{P}^{(1,2)}_x (\hat{c}^\dagger_{x+1}\hat{c}_x+\mathrm{H.c.})\hat{P}^{(1,2)}_x \; ,
	\end{equation}
	with the local projectors $\hat{P}^{(1)}_x = 1-(\hat{n}_{x+2}-\hat{n}_{x-1})^2$, which projects onto states where $\hat n_{x-1}=\hat n_{x+2}$, and $\hat{P}^{(2)}_x~=~1~+~\frac{1}{4}[(\hat{n}_{x-2}+\hat{n}_{x+2}-\hat{n}_{x-1}-\hat{n}_{x+3})^4 - 5 (\hat{n}_{x-2}+\hat{n}_{x+2}-\hat{n}_{x-1}-\hat{n}_{x+3})^2 ]$, which projects onto states with $n_{x-2}+n_{x+2}=n_{x-1}+n_{x+3}$, respectively.
	
	In the limit $V_1\to \infty$ and $V_2\to\infty$, both constrains act simultaneously\footnote{Here, we assume that we take the limits $V_1$ and $V_2$ \emph{independently} to define the Hamiltonian $\hat{H}^{(3)}_{\infty}$. If we took the limit $V\to\infty$ with $V_1=\alpha V$ and $V_2=\beta V$, we would project onto the eigenspaces of $\sum_{x}(\alpha \hat{n}_x\hat{n}_{x+1}+\beta \hat{n}_x\hat{n}_{x+2})$.}:
	\begin{equation}
		\hat{H}^{(3)}_\infty = - \sum_x \hat{P}^{(2)}_x\hat{P}^{(1)}_x (\hat{c}^\dagger_{x+1}\hat{c}_x+\mathrm{H.c.})\hat{P}^{(1)}_x\hat{P}^{(2)}_x 
	\end{equation}
	Note that $\hat{P}_x^{(1)}$ and $\hat{P}_x^{(2)}$ commute.
	
	There are three relevant operators that are preserved in certain limits, namely, the total number of particles $\hat{N}=\sum_x \hat{c}^\dagger_{x}\hat{c}_x $; the number of \textit{bonds} $\hat{N}^{(1)}_{\bullet \bullet} = \sum_x \hat{n}_{x}\hat{n}_{x+1}$, i.e. lattice bonds with both of the adjacent sites occupied; and the number of second-order bonds $\hat{N}^{(2)}_{\bullet\bullet}=\sum_{x}\hat{n}_x\hat{n}_{x+2}$.
	
	The Hamiltonian always conserves $N$, which allows us to focus on the half-filling sector ($N=\frac{L}{2}$) with dimension
	\begin{align}
		\mathcal{D}_{\mathrm{half}}=\binom{L}{\frac{L}{2}}\sim\sqrt{\frac{2}{L\pi}}2^L\,. 
	\end{align}
	In contrast, $N^{(1)}_{\bullet \bullet}$ is conserved only in the limit $V_1\to\infty$, and $N^{(2)}_{\bullet\bullet}$ is conserved in the limit $V_2\to\infty$. Thus, the Hilbert space of $\hat{H}^{(i)}_{\infty}$ fragments into disjoint blocks of product states characterized by their filling $N$ and respective bond numbers $N^{(1)}_{\bullet \bullet}$ and/or $N^{(2)}_{\bullet\bullet}$. As derived in the Supplemental Material~(part A), for a chain of $L$ sites the largest symmetry sector is characterized by $N=\frac{L}{2}$ and, depending on which quantities are conserved, $N^{(1)}_{\bullet \bullet}=\frac{L}{4}$ and/or $N^{(2)}_{\bullet\bullet}=\frac{L}{4}$. The dimension of these symmetry sectors scales with the dimension of the full Hilbert space, i.e., $2^L$, multiplied by some polynomial correction (see Supplemental Material, part A and Table~\ref{tab:blocks}).

	\textbf{Mapping to spins and movers.} It turns out that even within each symmetry sector there is further fragmentation due to the local constraints on particle hopping. In particular, \emph{frozen states} are product states that are simultaneously eigenstates of $\hat{H}^{(i)}_{\infty}$, so they correspond to one-dimensional blocks of $\hat{H}^{(i)}_{\infty}$. Let us illustrate this for $\hat{H}^{(1)}_{\infty}$ with $L=8$, $N=4$, and $N^{(1)}_{\bullet \bullet}=2$. Indicating filled (empty) by $\bullet$ ($\circ$), the state $\ket{ \bullet \! \bullet \! \circ \! \circ \! \bullet \! \bullet \! \circ  \circ}$ is frozen, as it is an eigenstate. Spatial translations generate four such states in total. One may use these states as a starting point to construct further blocks of states by shifting particles in a systematic fashion. The state $\ket{ \bullet \! \bullet \! \bullet \! \circ \! \circ \! \bullet \! \circ \circ}$, for example, is in the same symmetry sector but is clearly dynamically disconnected from any frozen state. The single particle on the right can hop freely until it encounters the block of three particles on the left, in which case it can assist in the hopping of said particles, i.e.,
	\begin{align}
		\begin{split}\label{eq:split}
			\ket{ \bullet \! \bullet \! \bullet \! \circ \! \circ \! \bullet \! \circ  \circ} &\rightarrow \ket{ \bullet \! \bullet \! \bullet \! \circ \! \bullet \! \circ \! \circ  \circ} \rightarrow \ket{ \bullet \! \bullet \! \circ \! \bullet \! \bullet \! \circ \! \circ \circ}\\
			&\rightarrow \ket{ \bullet \! \circ \! \bullet \! \bullet \! \bullet \! \circ \! \circ \circ} \rightarrow \ket{ \circ \! \circ \! \bullet \! \bullet \! \bullet \! \circ \! \circ \bullet} \rightarrow \dots
		\end{split}
	\end{align}
	The free particle is called a \textit{mover} for obvious reasons and the above example illustrates that a mover passing through the entire chain results in a state that is shifted by two sites with respect to the initial state; that is, in this case we would not have localization.
	
	In order to analyze the Hilbert space fragmentation in the limits $V_1\to\infty$ and $V_2\to\infty$, it is useful to map the states from the \emph{occupation number picture} to a \emph{spin/mover} picture\,\cite{Dias_2000,de2019dynamics}. For this, we adopt the following rules: Every bond ${\bullet\bullet}$ is assigned an up spin~$\muparrow$, and every pair of unoccupied sites ${\circ\circ}$ is assigned a down spin~$\mdownarrow$. Furthermore, we introduce so-called \emph{movers} $\mathbf{0}$ associated with alternating sequences, \eg, ${\bullet\!\circ\!\bullet}$, ${\circ\!\bullet\!\circ}$, ${\bullet\!\circ\!\bullet\circ}$, ${\circ\!\bullet\!\circ\bullet}$, etc., of length $n$, which is mapped to $\lfloor\tfrac{n-1}{2}\rfloor$ movers.
	
	This mapping can be compactly summarized as
	\begin{align}
		\ket{\underbrace{\bullet\!\bullet\!\cdots\!\bullet\!\bullet}_{n_1}\hspace{-2.33mm}\overbrace{\bullet\!\circ\!\bullet\!\cdots\!\circ\!\bullet\circ}^{n_2}\hspace{-2.33mm}\underbrace{\circ\!\circ\dots\circ\!\circ}_{n_3}}\,\,\equiv\,\, \ket{\underbrace{\muparrow\!\cdots\!\muparrow}_{n_1-1}\underbrace{\textbf{0}\!\cdots\!\textbf{0}}_{\lfloor\tfrac{n_2-1}{2}\rfloor}\underbrace{\mdownarrow\!\cdots\!\mdownarrow}_{n_3-1}},\label{eq:mapping-illustration}
	\end{align}
	where we illustrate two blocks of spins separated by a block of movers. Note that our counting of the mover sequences (length of alternating sequence) is such that the last and first sites also count towards the spin blocks [see overlapping braces in~\eqref{eq:mapping-illustration}]. For periodic boundary conditions the mapping to spin/movers is not injective, as sometimes two quantum states are mapped to the same spin/mover states (see the Supplemental Material, part B).
	
	\textbf{Effective dynamics.} In order to analyze fragmentation and localization in each of the three limits $\hat{H}^{(1,2,3)}_\infty$, we employ the general strategy of deriving effective hopping rules in the spin-mover picture. Using these rules, we systematically construct frozen states and large blocks and derive the asymptotic scaling of their number and dimension, respectively. In the case of $\hat{H}^{(1)}_\infty$ the effective rules are especially simple and it is clear that any product state but a frozen state dynamically delocalizes. The cases $\hat{H}^{(2,3)}_\infty$ require a more detailed analysis. We use statistical arguments and numerical evidence to show that the dynamics of $\hat{H}^{(2)}_\infty$ is also delocalized, while $\hat{H}^{(3)}_\infty$ is typically localized. However, introducing the average mover density as a control parameter gives rise to a transition between localized and delocalized dynamics in $\hat{H}^{(3)}_\infty$. 
	
	\textbf{Fragmentation for $V_1\to\infty$.} We introduce the number of spin ups $N_{\muparrow}$, the number of spin downs $N_{\mdownarrow}$, the number of movers $N_{0}$, and the number of times the pattern $\mdownarrow \muparrow$ appears in the spin sequence $N_{\mdownuparrows}$. Then we find the following constraints: $N_{\muparrow}=N_{\mdownarrow} = \frac{N}{2}$ and $N_{\mdownuparrows} + N_0 = N^{(1)}_{\bullet \bullet}$. Every state corresponds to a sequence of $\muparrow$, $\mdownarrow$, and $\textbf{0}$'s, and the constrained dynamics corresponds to the $\textbf{0}$'s moving freely through the sequence with the additional condition that moving every $\mathbf{0}$ through the entire chain results in a cyclic permutation of the spin sequence (PBCs imposed), i.e.,
	\begin{align}
		\begin{split}
			\ket{\muparrow\muparrow\mdownarrow\!\textbf{0}\!\mdownarrow}&\to\ket{\muparrow\muparrow\!\textbf{0}\!\mdownarrow\mdownarrow}\to\ket{\muparrow\!\textbf{0}\!\muparrow\mdownarrow\mdownarrow}\\
			&\to\ket{\textbf{0}\!\muparrow\muparrow\mdownarrow\mdownarrow}\to\ket{\mdownarrow\muparrow\muparrow\mdownarrow\mdownarrow\!\textbf{0}}\to\dots\,,
		\end{split}
	\end{align}
	which corresponds to~\eqref{eq:split} rewritten in spins. The dimension $\mathcal{D}_{\mathrm{frozen}}$ of the frozen state block for $V_1\to\infty$ and half filling can be computed analytically. For this, we recall that for $\hat{H}^{(1)}_{\infty}$ only the movers can move, so a state is frozen if it does not contain any movers, i.e., there are no $\mathbf{0}$'s in the spin picture. We can organize such frozen states by the number $2w$ of spin domains (up and down spins). For $L$ sites and $2w$ domains, there must be $\frac{L}{2}-w$ spins of each type, which we need to distribute over the different domains. Using the saddle point method reviewed in the Supplemental Material~(part C1), we find the asymptotics
	\begin{align}
		\mathcal{D}_\textnormal{frozen}\sim \sqrt{\tfrac{3\sqrt{5}-5}{\pi L}}\left(\tfrac{1+\sqrt{5}}{2}\right)^L=\frac{0.74}{\sqrt{L}}(1.62)^L\,.\label{eq:DforfrozenV1}
	\end{align}
	The number of frozen states is exponentially suppressed in the sector of half filling with $\mathcal{D}_{\mathrm{half}}=\binom{L}{\frac{L}{2}}\sim \sqrt{\frac{2}{\pi L}}\,2^L$. While their dynamics is trivially localized, a generic initial state will thus not have sufficient overlap with frozen states to make its dynamics also local.
	
	The largest block for $V_1\to\infty$ is found in\,\cite{de2019dynamics} using the spin/mover picture. It corresponds to states with $N_{0}=\frac{L}{4}-1$ movers (corresponding to $\frac{L}{2}$ sites) that move freely through two regions (one filled, one empty) of $\frac{L}{4}$ sites each, i.e., generated by
	\begin{align}
		\hspace{-1mm}\ket{{\bullet\!\circ\dots\bullet\!\circ\!\bullet\!\bullet\dots\bullet\!\bullet\!\circ\!\circ\dots\circ\!\circ}}=\ket{\mathbf{0}\cdots\mathbf{0}\!\muparrow\!\dots\!\muparrow\mdownarrow\!\dots\!\mdownarrow}\,.
	\end{align}
	Starting with the spin state $\ket{{\muparrow\!\dots\!\muparrow\mdownarrow\!\dots\!\mdownarrow}}$, we can cyclically rotate the spins to get $\frac{L}{2}$ different states, which can then be filled by distributing the $N_{0}$ movers over the $\frac{L}{2}$ positions between the spins. We therefore count a block dimension of
	\begin{align}
		\begin{split}
			\textstyle\mathcal{D}_{\max}=\frac{L}{2}\binom{\frac{L}{2}\!+\!N_{0}\!-\!1}{N_{0}}&\sim \sqrt{\tfrac{L}{27\pi}}\left(\tfrac{3^{3/4}}{\sqrt{2}}\right)^L=0.11\sqrt{L}(1.61)^L\,.
		\end{split}
	\end{align}
	We have $\lim_{L\to\infty}\frac{\mathcal{D}_{\max}}{\mathcal{D}_{\mathrm{sym}}}=0$, where $\mathcal{D}_{\mathrm{sym}}$ is the dimension of the relevant symmetry sector (see the Supplemental Material, part A), so $\hat{H}^{(1)}_\infty$ admits strong Hilbert space fragmentation. At the same time, we see explicitly that the movers $\mathbf{0}$ can move freely through the whole system. As this applies to all blocks, except frozen states, $\hat{H}^{(1)}_\infty$ will not show any localization for generic initial states.
	
	
	\begin{figure}
        \centering
	\includegraphics[width=\linewidth]{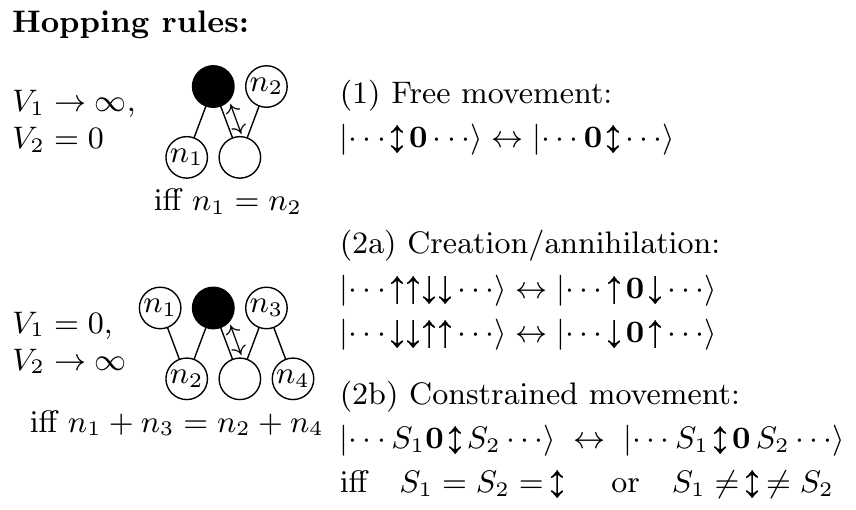}
		\caption{We illustrate the spin mapping, which allows us to map the fermionic Fock states to a sequence of spins $\mdownuparrows$ and so-called \emph{movers} $\mathbf{0}$ based on\,\cite{Dias_2000}. Movers move freely and are preserved in $\hat{H}^{(1)}_{\infty}$, while in $\hat{H}^{(2)}_{\infty}$ they are constrained and can be created/annihilated based on the illustrated rules. For $V_1,V_2\to\infty$, we have only the constrained movement (2b). The symbol $\mupdownarrow$ represents a spin that is fixed to either up or down.}
		\label{fig:rules}
	\end{figure}
	
	\textbf{Fragmentation for $V_2\to\infty$.} In order to simplify the dynamics and to bring similar combinatorial arguments to bear on this problem, we use the same mapping as before, i.e., every state is now written as a sequence of $\muparrow$, $\mdownarrow$, and $\textbf{0}$'s. However, the hopping rules are now given by rules (2a) and (2b) in Fig.~\ref{fig:rules}, where $\mupdownarrow$ represents either an up or down spin and $S_{1,2}\in\{\muparrow,\mdownarrow,\mathbf{0}\}$ can be any symbol. We see that it is still the movers ($\mathbf{0}$'s) that hop, although with further constraints. However, the number of movers is no longer conserved since they can be created and annihilated by certain spin configurations.
	
	Note that rule (2a) leads to a second-order hopping process for movers in certain configurations, e.g.,
	\begin{equation}
		\hspace{0mm}\ket{\cdots \!\muparrow \! \mathbf{0} \! \mdownarrow \muparrow \muparrow\! \cdots} \leftrightarrow \ket{\cdots\! \muparrow \muparrow \mdownarrow \mdownarrow \muparrow \muparrow \!\cdots} \leftrightarrow \ket{\cdots\! \muparrow \muparrow \mdownarrow \! \mathbf{0} \! \muparrow\! \cdots}.\label{eq:second-order-process}
	\end{equation}
	Since the model is symmetric under a global transformation $n_x \leftrightarrow 1-n_x$, there is a corresponding hopping process when globally substituting $\muparrow \, \leftrightarrow \, \mdownarrow$. Hence, we see that movers can trade position with antialigned spins if the pattern is framed by identical spins that are aligned with the outer spin of the antialigned pair, as illustrated in~\eqref{eq:second-order-process}.
	
	In the Supplemental Material~(part C2), we systematically construct frozen states as spin configurations that cannot generate movers and derive a lower asymptotic bound $\mathcal{D}_{\mathrm{frozen}}$ for the number of frozen states, namely,
	\begin{align}
		\mathcal{D}_{\mathrm{frozen}}\gtrsim \frac{4.98}{L}\,(1.22)^L\,.\label{eq:Nfronzen-V2}
	\end{align}
	
	By comparison with numerical results, it is clear that these spin states do not contribute the bulk of frozen states for $V_2 \to \infty$. Instead, states containing frozen movers represent the majority of such states in the asymptotic limit.
	
	In the Supplemental Material~(part C2), we construct a lower bound $\mathcal{D}^-_{\max}$ for the dimension of the largest block as some involved sum over binomial factors. We then expand the summand for large $L$, convert the sum into an integral and then extract the asymptotics using the saddle point method to find
	\begin{align}
		\mathcal{D}_{\max}\gtrsim\frac{4}{\sqrt{\pi L}}\sqrt[4]{5}^L\approx\frac{2.26}{\sqrt{L}}(1.50)^L\,.
	\end{align}
	
	\textbf{Fragmentation for $V_1,V_2\to\infty$.} The case where both interactions terms are taken \emph{independently} to infinity, i.e., $V_1\to\infty$ and $V_2\to\infty$, is characterized by the conservation of $\hat{N}$, $\hat{N}^{(1)}_{\bullet\bullet}$, and $\hat{N}^{(2)}_{\bullet\bullet}$. Therefore, rule (2b) from fig.~\ref{fig:rules} describes the only allowed hopping for movers.
	
	\begin{table}[t!]
		\centering
		\renewcommand{\arraystretch}{1.4}
		\begin{tabular}{c||c||c|c}
			&  $\mathcal{D}_{\mathrm{frozen}}$ & $\mathcal{D}_{\max}$ & $\mathcal{D}_{\mathrm{sym}}$\\
			\hline
			$V_1\to\infty$    & $\sim\frac{0.74}{\sqrt{L}} (1.62)^L$ & $\sim0.11\sqrt{L}\,(1.61)^L$ & $\sim\frac{1.27}{L} 2^L$\\
			$V_2\to\infty$ & $\gtrsim\frac{4.98}{L} (1.22)^L$ & $\gtrsim\frac{2.26}{\sqrt{L}} (1.50)^L$ & $\sim \frac{1.27}{L} 2^L$\\
			$V_1,V_2\to\infty$ & $\gtrsim\frac{0.74}{\sqrt{L}} (1.62)^L$  & $\gtrsim\frac{1.13}{\sqrt{L}} (1.41)^L$ & $\sim \frac{2.03}{L^{3/2}} 2^L$
		\end{tabular}
		\caption{We consider the half-filling sector $N=\frac{L}{2}$ and list the lower bounds $\mathcal{D}_{\mathrm{frozen}}$, for the number of frozen states in this sector, and $\mathcal{D}_{\max}$, for the dimension of the largest block in this sector. We compare them to $\mathcal{D}_{\mathrm{sym}}$, which is the dimension of the largest symmetry sector, i.e., $N^{(1)}_{\bullet \bullet}=\frac{L}{4}$ for $V_1\to\infty$, $N^{(2)}_{\bullet\bullet}=\frac{L}{4}$ for $V_2\to\infty$, and both combined for $V_1,V_2\to\infty$. While the symbol $\sim$ indicates exact asymptotics, the symbol $\gtrsim$ indicates only a lower bound for the asymptotics.
		}
		\label{tab:blocks}
	\end{table}
	
\begin{figure}[t!]
		\centering
		\includegraphics[width=\linewidth]{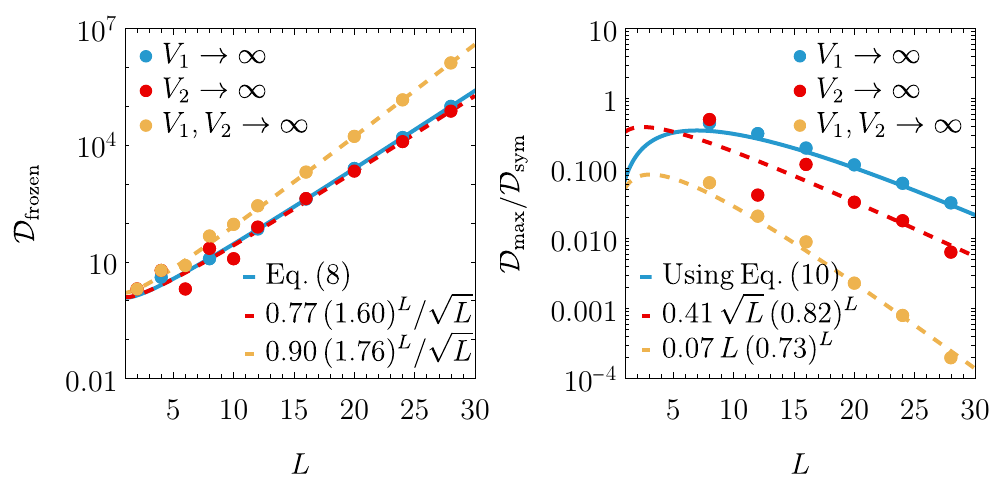}
		\caption{Analysis of  $\mathcal{D}_{\mathrm{frozen}}$ and $\mathcal{D}_{\max}/\mathcal{D}_{\mathrm{sym}}$ by comparing the exact values with our analytics ($V_1\to\infty)$ and numerical fits ($V_2\to\infty$, $V_1,V_2\to\infty$), confirming $\mathcal{D}_{\max}/\mathcal{D}_{\mathrm{sym}}\to 0$ as $L\to\infty$ (strong fragmentation). Here, $\mathcal{D}_{\max}$ refers to the largest Block within the largest symmetry sector, whose dimension is $\mathcal{D}_{\mathrm{sym}}$ (see Table~\ref{tab:blocks}).}
		\label{fig:scaling}
	\end{figure}

	All frozen states in the limit $V_1\to\infty$ are also frozen in the limit $V_1,V_2\to\infty$, so we immediately find the previously derived asymptotics~\eqref{eq:DforfrozenV1} for $V_1\to\infty$ as a lower bound, i.e.,
	\begin{align}
		\mathcal{D}_{\mathrm{frozen}}\gtrsim\sqrt{\tfrac{3\sqrt{5}-5}{\pi L}}\left(\tfrac{1+\sqrt{5}}{2}\right)^L=\frac{0.74}{\sqrt{L}}(1.62)^L\,.
	\end{align}
	From fig.~\ref{fig:scaling}, we see that this already describes the asymptotics well, but a numerical fit indicates that the leading coefficient should be around $2.59$ instead of $0.74$. We thus see that even in the limit $V_1,V_2\to\infty$ the frozen block still represents a fraction $\mathcal{D}_{\mathrm{frozen}}/\mathcal{D}_{\mathrm{half}}\to 0$ that vanishes exponentially fast, so that we need to focus on the larger blocks to analyze potential localization.
	
	For the construction of large blocks the same starting point can be used as in the case $V_2\to\infty$, since the contribution from states with $n=2 \lceil \frac{L}{8} \rceil$ and $N_\mathbf{0}=\frac{L}{2} - 2 \lceil \frac{L}{8} \rceil$ makes use of only rule (2b). We therefore find the lower bound on the dimension $\mathcal{D}_{\max}$ of the largest block to be
	\begin{align}
		\hspace{-2mm}\mathcal{D}_\textnormal{max} \geq \binom{n+N_\mathbf{0}}{N_\mathbf{0}} \!\sim\! \tfrac{4^{n}}{\sqrt{\pi n}}\!=\!\tfrac{\sqrt{2}^{L}}{\sqrt{\pi L/4}}\!\approx\!\tfrac{1.13}{\sqrt{L}}(1.41)^L\,.
	\end{align}
	
	\textbf{The problem of localization.} Our key findings are summarized in Table~\ref{tab:blocks} and fig.~\ref{fig:scaling}, where we study the dimensions of the frozen state space $\mathcal{D}_{\mathrm{frozen}}$ and of the largest block $\mathcal{D}_{\max}$ within the half-filling sector. We find in all three limits $V_1\to\infty$, $V_2\to\infty$, and $V_1,V_2\to\infty$ that the frozen states are exponentially suppressed ($\mathcal{D}_{\mathrm{frozen}}/\mathcal{D}_{\mathrm{half}}\sim e^{-\alpha L}$) and hence should not cause typical states to stay localized. We then study larger blocks and determine the dimension scaling of the largest block $\mathcal{D}_{\max}$. While we find freely moving $\mathbf{0}$s in the states of large blocks and thus delocalized dynamics, we also see that for $V_2\to\infty$ and $V_1,V_2\to\infty$ there exist Hamiltonian blocks generated from states where individual movers are constrained due to barriers that do not let movers pass through according to rule (2b) from fig.~\ref{fig:rules}. Thus, we conclude that $V_1\to\infty$ does not lead to localization, while the limits $V_2\to\infty$ and $V_1,V_2\to\infty$ require a more detailed analysis.
	
	We have seen that the limit $V_1\to\infty$ yields fragmentation, while the dynamics is not localized in real space (with the exception of frozen states). We may attribute the existence of large blocks to the fact that each mover is spatially unconstrained, so a finite number of movers contributes a binomial coefficient to the dimension of the block generated by such states. We were able to construct large blocks in the limit $V_2\to\infty$ by carefully choosing a generating state that allows for fully delocalized dynamics despite the restricted hopping rule (2b). This begs the question of whether full spatial delocalization is typical even for the cases $V_2\to\infty$ and $V_1,V_2\to\infty$. We will analyze this question using classical Markov simulations\,\cite{PhysRevB.101.214205}.
	
	\textbf{Markov simulations for $V_2\to\infty$.} We consider random states with a fixed density of movers $n_0$. We do this by first generating strings of spin and mover symbols of length $m$ and a total of $n_0 m$ movers. We subsequently convert these to states in real space characterized by their local occupancies. These states are then time evolved via a Markov simulation (see the Supplemental Material, part D), during which we keep track of active sites, i.e., those sites that change their occupancy at any point in the simulation. We find that essentially, all initial states result in a fully delocalized dynamics, except for very low densities $n_0$, as one might expect from the fact that large blocks are prevalent for $V_2 \to \infty$. We can explain this fact by showing explicitly that typical states should delocalize based on rules (2a) and (2b). The problem can be broken down into two parts. First, consider random strings of spins with no movers. Rule (2a) will generate a mover for any $\muparrow \muparrow \mdownarrow \mdownarrow / \mdownarrow \mdownarrow \muparrow \muparrow$ pattern within the string. These occur with a rate $r=\frac{N_{\muparrow \muparrow \mdownarrow \mdownarrow}}{m} = 1/8$ and should follow a Poisson distribution. Therefore, the distance between two consecutive patterns of this type should be Poisson distributed with mean $\mu=1/r=8$. Applying rule (2a) will reduce the number of symbols between the resulting movers by two, and we find that the symbol distance $\Delta$ between movers is a shifted Poisson distribution. We are left with a string of random spins that does not contain any $\muparrow \muparrow \mdownarrow \mdownarrow / \mdownarrow \mdownarrow \muparrow \muparrow$ patterns and a finite number of movers embedded within it. Second, we consider one such mover and its surrounding spins and study the size of the active region generated by it. It is clear that any activity has to spread locally from this single mover in a random spin background. We perform Markov simulations on random symbol strings of this type. For a given number $m$ of spin symbols to the right of the central mover we identify the site in real space that corresponds to the end of the $m$th symbol and keep track of its activity. This allows us to sample the cumulative probability distribution [$\mathrm{CDF}(m)$] of activity over a range of symbols $0 \leq m \leq 15$ (see Fig.S3 in the Supplemental Material). We find that the distribution saturates at a value of $\mathrm{CDF}_\mathrm{max}\approx 0.53$ at $m \approx 5$, indicating that in around half of all cases the active region is confined to fewer than five symbols and otherwise spreads indefinitely due to an avalanche effect. We can now find the probability that two consecutive movers produce active regions that do \textit{not} connect:
	
	\begin{equation}
		\begin{aligned}
			\mathcal{P}_{\mathrm{disc}} &= \sum_{\Delta} \mathcal{P}_\mathrm{sep}(\Delta) \sum_{m=0}^{\Delta-1} \mathrm{CDF}(m) \; \mathrm{PDF}(\Delta - m)\approx 0.26\,.
		\end{aligned}
	\end{equation}
	Here, $P_\mathrm{sep}$ is the distribution of distance between consecutive movers, and $\mathrm{PDF}$ is the probability density function associated with the cumulative distribution $\mathrm{CDF}$. The avalanche effect implies that essentially, every active region associated with a mover has to remain finite in order to avoid full delocalization and therefore the probability of localization for a random symbol string of length $m$ is $\mathcal{P}_{\mathrm{loc}} \approx (\mathrm{CDF}_\mathrm{max})^{mr}$. Adding additional movers by setting $n_0 > 0$ will affect $\mathcal{P}_\mathrm{sep}$ and hence lower the value of $\mathcal{P}_{\mathrm{disc}}$. However, its main effect on the probability of localization is via a shift in the exponent: $\mathcal{P}^{(n_0)}_{\mathrm{loc}} \approx (\mathrm{CDF}_\mathrm{max})^{m(r+n_0)}$. This explains why only very low initial mover densities $n_0$ and short chains yield a significant fraction of localized states.
	
	\textbf{Markov simulations for $V_1,V_2\to\infty$.} This case is characterized by only rule (2b), and the treatment in terms of active and inactive regions can be done analytically. A mover in a random spin background can hop past a given spin with probability $p=\frac{1}{2}$ according to (2b). Hence, the probability of creating an active region that extends over $m$ spins is
	\begin{equation}
		P_\mathrm{S}(m) = \sum_{n=0}^m p^n p^{n-m} (1-p)^2 = (m+1) \; 2^{-(m+2)}    .
	\end{equation}
	The mean free path of a mover is given by $\overline{m}~=~\sum_m m P_\mathrm{S}(m)~=~2$. Numerically, we find a compatible distribution in the real-space length $L$ using a Markov simulation with $t=10^5$ time steps and a sample of $1000$ initial symbol strings if we account for the relation $L \approx 1.65 m$.
	
	\begin{figure}[t!]
		\centering
		\includegraphics[scale=0.52]{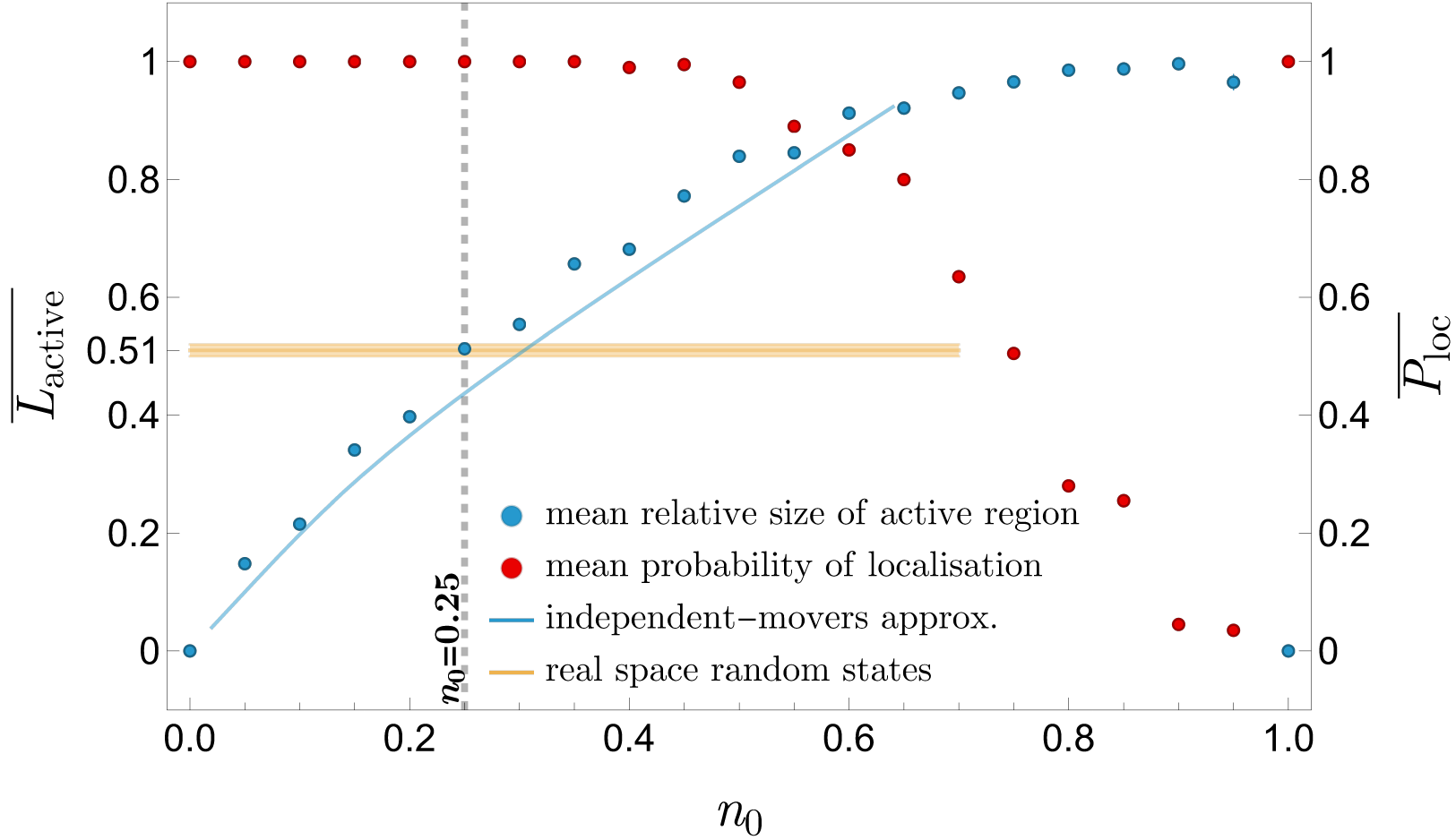}
		\caption{
		For $V_1,V_2\to\infty$, we show the mean size of the active regions $\overline{L_\mathrm{active}}$ relative to the total system size for random symbol strings with $M_\mathrm{total}=31$ symbols and $n_0 M_\mathrm{total}$ movers in blue.} The solid blue line corresponds to the predicted value of $\overline{L_\mathrm{active}}$ based on an independent-mover model. We show the mean active size for random states in real space with fixed length $L_\mathrm{total}=42$ in solid orange and the mean probability of localization $\overline{P_\mathrm{loc}}$ as a function of mover density for random symbol strings with $M_\mathrm{total}=31$ in red. The dotted gray line marks the intersection of blue and solid orange lines and hence indicates that real-space random states correspond to a mover density $n_0=0.25$. All Markov simulations use 400 random initial strings and $8\times10^4$ time steps each.
		\label{fig:activeSizeV12}
	\end{figure}
	
	In order to estimate the size of active regions as a function of mover density we make the ansatz
	\begin{equation}
		\frac{M_\mathrm{active}}{M_\mathrm{total}} = \frac{L_\mathrm{active}}{L_\mathrm{total}} = n_0 (\overline{m} - \overline{x}),
	\end{equation}
	where $M_\mathrm{active}$ and $L_\mathrm{active}$ are the number of active sites in symbol space and real space, respectively. Likewise, $M_\mathrm{total}$ and $L_\mathrm{total}$ are the total numbers of symbols and sites, respectively. $\overline{x}$ is the average overlap between neighboring active regions. We estimate $\overline{x}$ assuming that neighboring active regions are dynamically independent, which turns out to be a good approximation for not too large mover densities.
	
	Figure~\ref{fig:activeSizeV12} shows data obtained from Markov simulations along with a prediction based on the independent-mover approximation. The latter fails to be accurate above $n_0\approx0.7$, where we numerically find essentially no dependence on $n_0$. The probability of localization, defined as a non-vanishing number of inactive sites, is essentially $P_\mathrm{loc}=1$ up to $n_0\approx0.5$. Comparing the mean size of active regions for random states in real space, i.e., equal probability of occupancy $0$ or $1$ on each site, with that of random states in symbol space of fixed mover density, we find that the mean mover density of the former is $\overline{n_0}=0.25$. Therefore, random states in real space are well within the localized regime, and the limit $V_1,V_2\to\infty$ shows localized dynamics for typical initial states.
	
	\textbf{Conclusion.} In summary, we studied the spinless extended Fermi-Hubbard chain in various limits of strong repulsion; the presence of any next-nearest-neighbor interactions breaks integrability. The model exhibits Hilbert space fragmentation for only nearest- or next-nearest-neighbor interactions as well as for both interaction terms combined. We derived effective hopping rules and construct frozen states, allowing us to derive lower bounds for the number of the highly localized frozen states and the dimension of the largest Hilbert space block. In contrast to the nonextended Fermi-Hubbard model, our results suggest that the extended model features interaction-induced localization.
	To substantiate these findings, we presented classical Markov simulations and indeed found localization provided that the density of initial movers is sufficiently low.
	While all our results have been derived for the fermionic model, due to its close relationship to the hard-core boson variant\,\cite{cazalilla2011one,leblond2021universality,bianchi2021volume}, the spin-1/2 XXZ chain\,\cite{orbach1958linear,walker1959antiferromagnetic} and the six-vertex model\,\cite{lieb1967residual} we expect our results apply to a large class of systems.
	
\begin{acknowledgements}
We thank D. Mikhail for discussions and help with evaluating $\mathcal{D}_{\mathrm{frozen}}$ and $\mathcal{D}_{\max}$ for large $L$. S.R. acknowledges support from the ARC through Grant No. FT180100211. L.H. gratefully acknowledges support from the Alexander von Humboldt Foundation.
\end{acknowledgements}
	
\bibliography{references}

\clearpage
	
\section*{Supplemental Material}

	\subsection{Dimension of symmetry sectors}\label{app:symmetry}
	We compute the dimension $\mathcal{D}_{\mathrm{sym}}$ of the largest symmetry sectors of $\hat{H}$ in the respective limits, \ie with fixed $N$, $N^{(1)}_{\bullet \bullet}$ and/or $N^{(2)}_{\bullet\bullet}$.
	
	\textbf{Limit $V_1\to\infty$.} The largest symmetry sector corresponds to $N=\frac{L}{2}$ and $N^{(1)}_{\bullet \bullet}=\frac{L}{4}$. To determine its dimension, we first consider $\frac{L}{2}$ filled sites arranged in a circle having $\frac{L}{2}$ bonds. We break $\frac{L}{4}$ of those by inserting empty sites, for which there are $\binom{L/2}{L/4}$ choices. We can now distribute an additional $L/4$ empty sites $\circ$, but only adjacent to existing empty sites to not break any further bonds, so that we have $\binom{L/2-1}{L/4}$ possibilities. Finally, there is an extra factor of $2$, as we can invert $\bullet\leftrightarrow\circ$ for each such state. The dimension of this maximal symmetry sector is thus
	\begin{align}
		\mathcal{D}_{\mathrm{sym}}
		=2\binom{\frac{L}{2}\!-\!1}{\frac{L}{4}}\binom{\frac{L}{2}}{\frac{L}{4}}=\binom{\frac{L}{2}}{\frac{L}{4}}^2\sim\frac{4}{\pi L} 2^L=\tfrac{1.27}{L} 2^L\,,\label{eq:DsymV1}
	\end{align}
	\ie the sector has the same exponential scaling as the entire Hilbert space with some polynomial correction.
	
	\textbf{Limit $V_2\to\infty$.} This case can be treated almost identically to the limit $V_1\to\infty$. For this, we rearrange our lattice sites $(1,2,3\dots,L)$ as $(1,3,\dots,L,2,4,\dots,L-1)$ assuming that $L$ is odd. Counting the next-nearest bonds $N^{(2)}_{\bullet \bullet}$ in the original order is now identical to counting the number of nearest bonds $N^{(1)}_{\bullet \bullet}$ in the rearranged lattice. As the distinction between odd and even number of lattice sites is subleading in the limit $L\to\infty$, we can therefore use our previous calculation for the nearest bonds $N^{(1)}_{\bullet \bullet}=\frac{L}{4}$, where $L$ was a multiple of $4$, and find again
	\begin{align}
		\mathcal{D}_{\mathrm{sym}}\sim\frac{4}{\pi L} 2^L=\tfrac{1.27}{L} 2^L\label{eq:DsymV2}
	\end{align}
	for the symmetry sector with $N=\frac{L}{2}$ and $N^{(2)}_{\bullet \bullet}=\frac{L}{4}$.
	
	\textbf{Limit $V_1,V_2\to\infty$.} We compute the dimensions of different symmetry sectors by assuming a random distribution of $\frac{L}{2}$ particles over $L$ sites (half filling). The probability that a given site is occupied is $p=\frac{1}{2}$ and therefore we find that the probability that a (next-)nearest-neighbor bond is formed with the next site is $p^{(1,2)}_{\bullet \bullet}=\frac{1}{4}$. However, bonds on consecutive sites are correlated and so are occupation number and bonds. The probability of finding $N^{(1,2)}_{\bullet \bullet}$ (next-)nearest neighbor bonds over $L$ sites therefore slightly deviates from a Binomial distribution with probability mass function (PMF)
	\begin{align}
		B(L,p^{(1,2)}_{\bullet \bullet})=\binom{L}{N^{(1,2)}_{\bullet \bullet}} \left(p^{(1,2)}_{\bullet \bullet}\right)^{N^{(1,2)}_{\bullet \bullet}} (1-p^{(1,2)}_{\bullet \bullet})^{L-N^{(1,2)}_{\bullet \bullet}}\,.
	\end{align}
	In the limit $L \to \infty$, we nevertheless approach a multivariate Gaussian distribution
	\begin{align}
		f_{\mathbf{n}}(N, N^{(1)}_{\bullet \bullet}, N^{(2)}_{\bullet \bullet})=\tfrac{\exp \left(-\frac{1}{2}(\mathbf{N}-\boldsymbol{\mu})^{\mathrm{T}} \boldsymbol{\Sigma}^{-1}(\mathbf{N}-\boldsymbol{\mu})\right)}{\sqrt{(2 \pi)^{3}\det\mathbf{\Sigma}}}\,,
	\end{align}
	where $\mathbf{N}=(N, N^{(1)}_{\bullet \bullet}, N^{(2)}_{\bullet \bullet})$. The expectation values are not affected by these local correlations and so we find $\mathbf{\mu}~=~(\langle N \rangle, \langle N^{(1)}_{\bullet \bullet} \rangle, \langle N^{(2)}_{\bullet \bullet} \rangle )~=~(\frac{1}{2},\frac{1}{4},\frac{1}{4})$. The covariance matrix $\mathbf{\Sigma}_{\alpha \beta}=\mathrm{Cov}(\mathbf{N}_\alpha,\mathbf{N}_\beta)=\braket{\mathbf{N}_\alpha\mathbf{N}_\beta}-\braket{\mathbf{N}_\alpha}\braket{\mathbf{N}_\beta}$ with $\mathrm{Var}(\mathbf{N}_\alpha)=\mathrm{Cov}(\mathbf{N}_\alpha,\mathbf{N}_\alpha)$ can be computed as follows:
	\begin{align}
		\begin{split}
			\mathrm{Var}(N) &= \sum_{x=1}^L \tfrac{1}{2} + \sum_{\substack{x,y = 1 \\ x \neq y}}^L \tfrac{1}{2} \cdot \tfrac{1}{2} - \left(\tfrac{L}{2}\right)^2 \\
			&= \tfrac{L}{2} + \tfrac{L(L-1)}{4} - \tfrac{L^2}{4} =\tfrac{L}{4}\,,
		\end{split}  \\
		\begin{split}\label{eq:varNbond}
			\mathrm{Var}(N^{(1)}_{\bullet \bullet})&= \sum_{x=1}^L \tfrac{1}{4} + 2 \sum_{x=1}^L \tfrac{1}{4} \cdot \tfrac{1}{2} + \sum_{\substack{x,y = 1 \\ |x-y|>1}}^L \tfrac{1}{4} \cdot \tfrac{1}{4} - \left(\tfrac{L}{4}\right)^2 \\
			&= \tfrac{L}{4} + \tfrac{2L}{8} + \tfrac{L(L-3)}{16} - \tfrac{L^2}{16} = \tfrac{5L}{16}\,,
		\end{split}     \\
		\begin{split}
			\mathrm{Cov}(N,N^{(1)}_{\bullet \bullet})&=  2 \sum_{x=1}^L \tfrac{1}{2} \cdot \tfrac{1}{2} + \sum_{\substack{x,y = 1 \\ x-1 \neq y \neq x}}^L \tfrac{1}{2} \cdot \frac{1}{4} - \tfrac{L}{2} \cdot \tfrac{L}{4} \\
			&= \tfrac{L}{2} + \tfrac{L(L-2)}{8} - \tfrac{L^2}{8} = \tfrac{L}{4}\,.
		\end{split}
	\end{align}
	By way of example we explicitly discuss the details of computing $\mathrm{Var}(N^{(1)}_{\bullet \bullet})$. First, we note that $N^{(1)}_{\bullet \bullet}=\sum_x n^{(1)}_{\bullet \bullet,x}$, where $n^{(1)}_{\bullet \bullet,x} = 1 (0)$ if site $x$ does (not) contribute a nearest neighbor bond, which is the case if and only if both site $x$ and $x+1$ are occupied. Hence we have to compute the double sum $\langle N^{(1)}_{\bullet \bullet} N^{(1)}_{\bullet \bullet} \rangle = \sum_{x,y} \langle n^{(1)}_{\bullet \bullet,x} n^{(1)}_{\bullet \bullet,y} \rangle$ where only those terms contribute for which both sites $x$ and $y$ form a bond with their nearest neighbor to the right. The first term in \ref{eq:varNbond} is the diagonal contribution from $x=y$. The second term represents the (identical) contributions from $y=x\pm1$, \ie the probability of finding a bond for site $x$ is $\frac{1}{4}$, but in the presence of such a bond the conditional probability of the neighboring sites $x\pm1$ contributing a bond is $\frac{1}{2}$. For sites that are further apart, \ie  $|x-y|>1$, the probabilities are independent and therefore $\frac{1}{4}$ each. This results in the third term in \ref{eq:varNbond}. Finally, we have to subtract $\langle N^{(1)}_{\bullet \bullet} \rangle^2 = (L/4)^2$. 
	
	The remaining variance $\mathrm{Var}(N^{(2)}_{\bullet \bullet})=\frac{5L}{16}$ and covariance $\mathrm{Cov}(N,N^{(2)}_{\bullet\bullet}) = \frac{L}{4}$ can be computed in exactly the same way as $\mathrm{Var}(N^{(1)}_{\bullet \bullet})$ and $\mathrm{Cov}(N,N^{(1)}_{\bullet \bullet})$, respectively. The dimension of the largest symmetry sector $(N=\frac{L}{2}, N^{(1)}_{\bullet \bullet}=\frac{L}{4}, N^{(2)}_{\bullet \bullet}=\frac{L}{4} )$ is asymptotically given by the dimension of the full Hilbert space, $\mathcal{D}=2^L$, times the maximum of the Gaussian distribution $f_\mathrm{max}$, \ie
	\begin{align}
		\mathcal{D}_{\mathrm{sym}}\sim\frac{2^L}{\sqrt{(2\pi)^3\det\mathbf{\Sigma}}}=\sqrt{\tfrac{128}{\pi^3L^3}}2^L=\tfrac{2.03}{L^{3/2}}2^L\,.\label{eq:DsymV12}
	\end{align}
	
	We compare our analytical derivation with the exact dimensions computed in figure~\ref{fig:Dsym-asymptotics}.
	
\renewcommand{\thefigure}{S1}	
	\begin{figure}[t!]
		\centering
		\includegraphics[width=.9\linewidth]{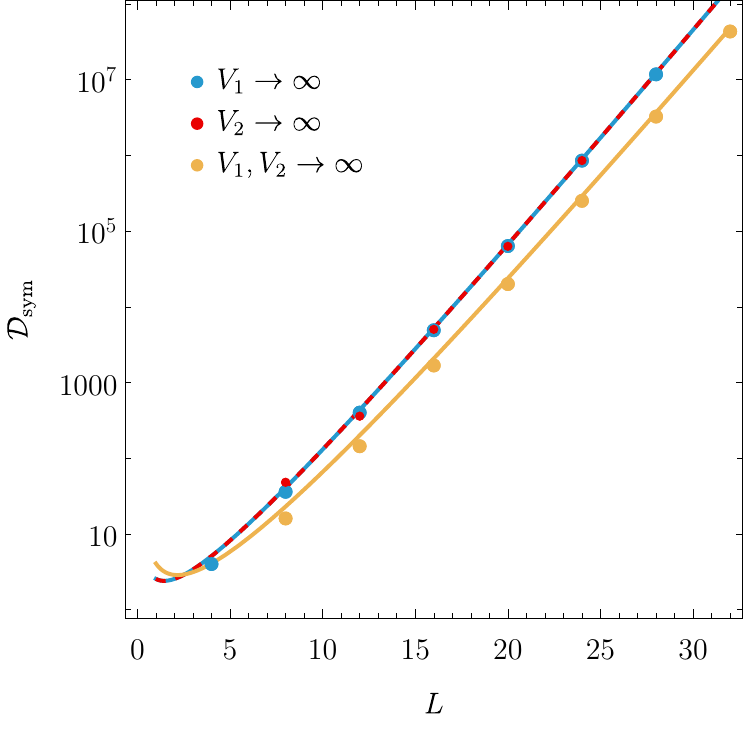}
		\caption{Comparison of the analytical asymptotics (curves) shown in~\eqref{eq:DsymV1},~\eqref{eq:DsymV2} and~\eqref{eq:DsymV12} with the exact values (data points).}
		\label{fig:Dsym-asymptotics}
	\end{figure}
	
	\subsection{Spin/mover picture and periodic boundary conditions}
	There is an important subtlety related to converting the occupation description ($\bullet$, $\circ$) into the spin/mover picture ($\muparrow$, $\mdownarrow$, $\mathbf{0}$) in the presence of periodic boundary conditions. This implies that in certain situations the map to spins/movers is not injective, \ie sometimes two different occupation states are mapped to the same spin/mover state. We will now explain how this degeneracy arises, which plays a role when counting the dimensions of different blocks in the Hamiltonians.
	
	We consider periodic boundary conditions and adopt the convention that spins representing $\bullet\bullet$ or $\circ\circ$ from the last to the first site appears at the end of our sequence, \ie $\ket{\circ\!\bullet\!\bullet\circ}\equiv\ket{\muparrow\mdownarrow}$. When the first $n$ sites are part of an uninterrupted mover sequence, we map them to $\lfloor\tfrac{n}{2}\rfloor$ movers, \eg $\ket{\circ\!\bullet\!\circ\!\circ\!\bullet}\equiv\ket{\mathbf{0}\!\mdownarrow\!\mathbf{0}}$ with $n=3$ and $\ket{\bullet\!\circ\!\circ\!\bullet\!\circ}\equiv\ket{\mdownarrow\!\mathbf{0}\mathbf{0}}$ with $n=2$, but $\ket{\circ\!\circ\!\bullet\!\circ\!\bullet}\equiv\ket{\mdownarrow\!\mathbf{0}\mathbf{0}}$ with $n=1$.
	
	With these conventions, every occupation state is mapped to a spin state, but this mapping is not one-to-one. Due to the periodic boundary conditions, certain spin/mover states correspond to two different states in the occupation picture (displaced by one site), whenever the last symbol is different from the first, \eg $\ket{\muparrow\mdownarrow}$ corresponds to both $\ket{\bullet\!\bullet\!\circ\circ}$ and $\ket{\circ\!\bullet\!\bullet\circ}$. Similarly, the state of only movers $\ket{\mathbf{0}\dots\mathbf{0}}$ corresponds to both $\ket{\bullet\!\circ\dots\bullet\!\circ}$ and $\ket{\circ\!\bullet\dots\circ\!\bullet}$. To build some further intuition consider the following examples:
	\begin{align}
		\begin{split}\label{eq:mapping-examples}
			\ket{\muparrow\mdownarrow\muparrow\downarrow}&\,\,\equiv\,\,\ket{\bullet\!\bullet\!\circ\!\circ\!\bullet\!\bullet\!\circ\circ}\,\text{or}\,\ket{\circ\!\bullet\!\bullet\!\circ\!\circ\!\bullet\!\bullet\circ}\,,\\
			\ket{\mathbf{0}\mathbf{0}\mathbf{0}\mathbf{0}}&\,\,\equiv\,\,\ket{\bullet\!\circ\!\bullet\!\circ\!\bullet\!\circ\!\bullet\circ}\,\text{or}\,\ket{\circ\!\bullet\!\circ\!\bullet\!\circ\!\bullet\!\circ\bullet}\,,\\
			\ket{\muparrow\muparrow\!\mathbf{0}\!\mdownarrow\mdownarrow}&\,\,\equiv\,\,\ket{\bullet\!\bullet\!\bullet\!\circ\!\bullet\!\circ\!\circ\circ}\,\text{or}\,\ket{\circ\!\bullet\!\bullet\!\bullet\!\circ\!\bullet\!\circ\circ}\,,\\
			\ket{\muparrow\!\mathbf{0}\mathbf{0}\!\mdownarrow}&\,\,\equiv\,\,\ket{\bullet\!\bullet\!\circ\!\bullet\!\circ\!\bullet\!\circ\circ}\,\text{or}\,\ket{\circ\!\bullet\!\bullet\!\circ\!\bullet\!\circ\!\bullet\circ}\,,\\
			\ket{\muparrow\!\mathbf{0}\mathbf{0}}&\,\,\equiv\,\,\ket{\bullet\!\bullet\!\circ\!\bullet\!\circ}\,\text{or}\,\ket{\circ\!\bullet\!\bullet\!\circ\!\bullet}\,,\\
			\ket{\muparrow\muparrow\muparrow}&\,\,\equiv\,\,\ket{\bullet\!\bullet\!\bullet}\,\,,\,\,\ket{\muparrow\mdownarrow\muparrow}\,\,\equiv\,\,\ket{\bullet\!\bullet\!\circ\!\circ\!\bullet}\,.
		\end{split}
	\end{align}
	The two-fold degeneracy will not affect the utility of the mapping, as we will use it specifically to analyze how the size of certain connected blocks in the Hamiltonian matrix scales. For this, an additional constant factor will not matter. In the case of $\hat{H}^{(1)}_\infty$, each of these two states will actually belong to separate blocks and therefore give rise to an overall factor of $2$ in the number of degenerate blocks without affecting their size. Only when counting the number of frozen states, we need to remember which states do not come with a degeneracy of $2$.

	\subsection{Combinatorics of sector dimensions}
	In the following, we will derive the scaling laws of the respective block sizes / number of frozen states listed in table~1 of the main text.
	
	\subsubsection{Frozen states for \texorpdfstring{$V_1\to\infty$}{V1->infty}}\label{app:frozen-V1}
	For a system with $L$ sites, we want to find the number of frozen states with half-filling, \ie $N=\frac{L}{2}$. In the spin picture, each such state
	\begin{align}
		\ket{\muparrow\!\cdots\!\muparrow\mdownarrow\!\cdots\!\mdownarrow\muparrow\!\cdots\!\muparrow\ldots}
	\end{align}
	is characterized by a sequence of spin domains. Periodicity requires that there must be an even number $2w$ of domains, \ie $w$ domains with spin up and $w$ domains with spin down. Recall that ${\muparrow\equiv\bullet\bullet}$, ${\muparrow\muparrow\equiv\bullet\!\bullet\!\bullet}$ etc., so a domain of $\ell$ spins corresponds to $\ell+1$ physical sites. If we have $2w$ domains with $2s$ spins (in total), we thus have $L=2s+2w$ sites. In order to represent a system with even $L$ sites, we thus must have a total of $s=\frac{L}{2}-w$ spins of each type ($2s$ in total). We therefore reduced the problem to counting all such spin configurations.
	
	For the purpose of counting, we can start with $s$ up spins arranged in a circle, which we need to break up at $w$ spots to place the $w$ domains of down spins. At this stage, we need to distinguish if we break between the last and the first spin or not. If we do not, there are $\binom{s-1}{w}$ ways to break up the circle and then an additional $\binom{(s-w)+w-1}{w-1}$ ways to distribute $s-w$ spin downs over the $w$ domains (each already containing one spin down). Each of these spin states corresponds to a unique occupancy configuration and we can finally flip $\bullet\leftrightarrow\circ$ to get a total degeneracy of $2$. If we break the circle between the last and the first spin, there are only $\binom{s-1}{w-1}$ such ways, while the counting of the possible distributions of down spins is still $\binom{(s-w)+w-1}{w-1}$. If we break the spin domain between the first and last spin, each spin configuration corresponds to two states in the occupancy picture (see explanation before~(8) of our main text) and we can again flip $\bullet\leftrightarrow\circ$ to get a total degeneracy of $4$. Finally, the states $\ket{\circ\!\bullet\!\dots\!\circ\!\bullet}$ and $\ket{\circ\!\bullet\!\dots\!\circ\!\bullet}$ both correspond to $\ket{\mathbf{0}\dots\mathbf{0}}$ in the spin picture and they are the only frozen states containing movers. The resulting sum
	\begin{align}
		\begin{split}
			\mathcal{D}_{\mathrm{frozen}}&\!=\!2\!+\!\sum^{L/4}_{w=1}\left(2\binom{s\!-\!1}{w}\binom{s\!-\!1}{w\!-\!1}\!+\!4\binom{s\!-\!1}{w\!-\!1}^2\right)\\
			&=2+\sum^{L/4}_{w=1}\frac{Lw(\frac{L}{2}-w-1)!^2}{(\frac{L}{2}-2w)!^2w!^2}
		\end{split}
	\end{align}
	can be calculated as hypergeometric function
	\begin{align}
		\begin{split}\label{eq:Dfrozen-hyper}
			&\mathcal{D}_{\mathrm{frozen}}\!=\\
			&2+{}_4F_3(1\!-\!\tfrac{L}{4},1\!-\!\tfrac{L}{4},\tfrac{3}{2}\!-\!\tfrac{L}{4},\tfrac{3}{2}\!-\!\tfrac{L}{4};2,2\!-\!\tfrac{L}{2},2\!-\!\tfrac{L}{2};16)\,L
		\end{split}
	\end{align}
	for even $L>2$. We can derive its asymptotics for $L\to\infty$ using the saddle point approximation. For this, we define $w=x L$, such that $x\in[0,\tfrac{1}{4}]$ and expand the summand for $L\to\infty$ as
	\begin{align}
		I(x)=\frac{L^2x(\frac{L}{2}-Lx-1)!^2}{(\frac{L}{2}-2Lx)!^2(Lx)!^2}\sim\frac{xe^2}{2\pi L} e^{f_1(x)L+f_2(x)}\,,
	\end{align}
	where the respective functions are
	\begin{align}
		f_1(x)&=\log \left(\tfrac{1\!-\!2 x}{1-4 x}\right)+2x\log\left(\tfrac{(1-4x)^2}{(2-4x)x}\right),\\
		f_2(x)&=\log \left(\tfrac{4 x}{8 x^2-6 x+1}\right)\,.
	\end{align}
	The maximum of $f_1(x)$ on $[0,\tfrac{1}{4}]$ is at $x_0=\frac{5-\sqrt{5}}{20}$, which allows us to expand $I(x)$ as
	\begin{align}
		I(x)\sim \frac{e^2x_0}{2\pi L}e^{(f_1(x_0)+ x^2f''_1(x_0)/2)L+f_2(x_0)}
	\end{align}
	around $x_0$. We then replace the sum by a Gaussian integral, \ie $\sum_w\approx L\int dx$, where we can extend the domain of the integral to the whole real line (as the integrand is exponentially surpressed away from $x_0$), yielding
	\begin{align}
		\begin{split}
			\mathcal{D}_{\mathrm{frozen}}&\sim L\int^{\infty}_{-\infty} dx I(x)=\tfrac{e^2x_0}{2\pi}\sqrt{\tfrac{\pi}{L |f''(x_0)|}}e^{f(x_0)L+f_2(x_0)}\\
			&=\sqrt{\tfrac{3\sqrt{5}-5}{\pi L}}\left(\tfrac{1+\sqrt{5}}{2}\right)^L=\frac{0.74}{\sqrt{L}}(1.62)^L\,.
		\end{split}
	\end{align}
	
	\subsubsection{Frozen states for \texorpdfstring{$V_2\to\infty$}{V2->infty}}\label{app:frozen-V2}
	In order to derive a lower bound on the number of frozen states, we can systematically construct such states starting from a Néel configuration of spins $\ket{\muparrow \mdownarrow \dots \muparrow \mdownarrow }$. We can now generate more frozen states by flipping one up-spin and one down-spin if the two are more than three sites apart in order to avoid $\muparrow \muparrow \mdownarrow \mdownarrow$ patterns, \eg
	\begin{equation}
		\ket{\cdots\! \muparrow \mdownarrow \muparrow \mdownarrow \muparrow \mdownarrow \muparrow \mdownarrow \muparrow \mdownarrow\! \cdots} \quad \rightarrow \quad \ket{\cdots\! \muparrow \mdownarrow \mdownarrow \mdownarrow \muparrow \mdownarrow \muparrow \muparrow \muparrow \mdownarrow\! \cdots}\,.
	\end{equation}
	However, this reduces the number of particle $N$ by $2$ and the number of sites $L$ (in the occupation picture) by $4$, which we can compensate for by adding another pair of $\muparrow \mdownarrow$ at the end of the chain:
	\begin{equation}
		\ket{\cdots\! \muparrow \mdownarrow \muparrow \mdownarrow \muparrow \mdownarrow \muparrow \mdownarrow \muparrow \mdownarrow \!\cdots} \quad \rightarrow \quad \ket{\cdots \!\muparrow \mdownarrow \mdownarrow \mdownarrow \muparrow \mdownarrow \muparrow \muparrow \muparrow \mdownarrow \!\cdots\! \muparrow \mdownarrow}\,.
	\end{equation}
	The previous steps can be repeated to flip more spins and will yield frozen states as long as the flipped spins are sufficiently separated. Since the number of available up-spins and down-spins scales as $\sim\frac{n}{2}$ we find that if we flip $k$ up-spins, we are left with $\sim \frac{n}{2}+k-4k$ down-spins available for flipping. Here, we assume that the flipped up-spins are sufficiently sparse, so that there are $2$ down-spins on either side of each flipped up-spin and recall that each flipped spin also adds $\muparrow \mdownarrow$. The number of states with $k$ flipped up- and down-spins thus scales at least as $\sim \binom{n/2}{k} \binom{n/2-3k}{k}$. With this, we find the asymptotics
	\begin{align}
		\binom{\frac{n}{2}}{xn}\binom{\frac{n}{2}-3xn}{xn}\sim \frac{1}{\pi n} e^{f_1(x)n+f_2(x)}\,,
	\end{align}
	where we have the functions
	\begin{align}
		\begin{split}
			f_1(x)&=\mathrm{arctanh}\left(\frac{2(1-5x)x}{1-10(1-x)x}\right)+x\log\tfrac{1-3x}{4x^2}\\
			&\quad -4x\log(1-8x)+5x\log(1-10x)\,,
		\end{split}\\
		f_2(x)&=\frac{1}{2}\log\left(\frac{1-8x}{4-16x(3-5x)}\right)-\log(x)\,.
	\end{align}
	As we are trying to find a maximal lower bound, we want to find the maximum of $f_1(x)$ by solving for $f'_1(x_0)=0$. This results in a transcendental equation, whose solution can be found numerically to be $x_0=0.077$. After replacing $n=\frac{L}{2}$, we then find the lower bound
	\begin{align}
		\mathcal{D}_{\mathrm{frozen}}\gtrsim\frac{4.98}{L}\,(1.22)^L\,.
	\end{align}
	
	\subsubsection{Large blocks for \texorpdfstring{$V_2\to\infty$}{V2->infty}}\label{app:large-V2}
	Now we want to provide a lower bound for the dimension of the largest block. Again, we systematically construct large connected blocks starting from a Néel state by adding $\mathbf{0}$s in between the spins such that no two $\mathbf{0}$s are adjacent. This allows for straightforward counting of the number of sites and particles, as adding a $\mathbf{0}$ in between two spins of an anti-aligned spin-pair, \ie $\ket{\cdots \muparrow \mdownarrow \cdots} \rightarrow \ket{\cdots \muparrow \! \mathbf{0} \! \mdownarrow \cdots}$, changes $N$ and $L$ in the same way as adding two spins: $\ket{\cdots \muparrow \mdownarrow \cdots} \rightarrow \ket{\cdots \muparrow \muparrow \mdownarrow \mdownarrow \cdots}$. The maximum number of $\mathbf{0}$s that can be added to a Néel state of length $n$ in this way is $n$. For any even chain of length $L$ we now find the shortest Néel state that gives the correct number of sites when adding an appropriate number $\leq n$ of $\mathbf{0}$s to it. In short, we have the conditions $L = 2(n+N_\mathbf{0})$, $N_\mathbf{0} \leq n$, $n$ even and we maximise $N_\mathbf{0}$. Therefore, we find $n=2 \lceil \frac{L}{8} \rceil$ and $N_\mathbf{0}=\frac{L}{2} - 2 \lceil \frac{L}{8} \rceil$. Both $n$ and $N_\mathbf{0}$ scale linearly in the system size $L$ and $n-N_\mathbf{0}<4$. For example, in the case $L=14$, we would find that at least four spins are required and three $\mathbf{0}$s need to be added. One such state might be $\ket{\muparrow \! \mathbf{0} \! \mdownarrow \! \mathbf{0} \! \muparrow \! \mathbf{0} \! \mdownarrow}$. A crucial observation is that the $\mathbf{0}$s can move freely throughout a Néel pattern, even in the presence of other movers. The hopping dynamics of our movers is therefore the same as in the case $V_1\to\infty$ with the exceptions of the additional possibility of conversion processes that annihilate one ore more of the movers and the second order hopping processes. We note that no additional movers can be created beyond the maximum number set by the construction outlined above and that the second order processes are only relevant after at least one of the initial movers has been annihilated.
	
	Based on the observation that the unrestricted hopping rules apply to a set of movers in a Néel background, we find that the block generated by this state contains at least $\binom{n+N_\mathbf{0}}{N_\mathbf{0}} \cdot 2$ states with $N_\mathbf{0}$ movers, where the factor of $2$ stems from cyclic permutation of the Néel pattern. Unlike the case $V_1 \rightarrow \infty$, we find numerically that states related through translation by one site may or may not be part of the same connected subspace. Therefore we have an additional factor of 2 in certain cases, \eg for $L=12$.
	
	Next, we consider states that involve less than the maximum number of movers. These can be generated by converting $\mathbf{0}$s into spins. Doing so will insert a $\ket{\cdots \muparrow \muparrow \mdownarrow \mdownarrow \cdots}$ (or flipped) pattern into the Néel state which acts as a barrier for the remaining $\mathbf{0}$s in regards to the first order hopping process. Ignoring second order hopping for now, we can generate every such state from one of the above states by converting the appropriate number of $\mathbf{0}$s to spins. It therefore becomes a simple task of counting the number of distinct states with maximal $N_\mathbf{0}$ that are characterised by exactly one $\mathbf{0}$ being separated from the others, two $\mathbf{0}$s being separated from every other, and so on. We define $\#(n,N_\mathbf{0},k)$ to be the number of distinct states with a total number of $n+N_\mathbf{0}$ symbols, $N_\mathbf{0}$ movers and $k < N_\mathbf{0}-1$ of these movers being mutually separated from each other and any groups of movers. A lower bound to this is given by the number of such states that only contain one connected group of movers with all other movers mutually separated, so $\#(n,N_\mathbf{0},k) \geq (n+N_\mathbf{0}) \binom{n-1}{k}$.
	
	More generally, we can study the case $n=N_\mathbf{0}$ (assuming $\frac{N}{8}\in\mathbb{N}$). Separating the movers into $0<X<N_\mathbf{0}$ groups of which there are $Y$ containing only a single mover, we have
	\begin{equation}
		\binom{x}{y} \cdot \binom{N_\mathbf{0}-X-1}{X-Y-1} \cdot \binom{n-1}{X-1} \cdot 2 \cdot 2
	\end{equation}
	such states. Here, we undercount because $\binom{N_\mathbf{0}-X-1}{X-Y-1}$ does not count states where group breaks up around boundary as distinct, but it suffices to get a lower bound. The first binomial coefficient is the number of ways to choose $Y$ single-mover groups out of $X$ total groups. The second one is the number of ways to distribute the $N_\mathbf{0}-X$ left over movers over the $X-Y$ groups with at least one mover added to each of those. The third binomial coefficient distributes $n$ spin symbols in between $x$ groups of movers, again adding at least one to each bin. Finally, there is one factor of 2 because the configuration can start with either a group of spins or a group of movers, and another factor of 2 because the Néel pattern can start with an up- or a down-spin. For fixed $Y$, we need to sum over $x$ with the constraint $Y<X \leq \frac{N_\mathbf{0}+Y}{2}$:
	\begin{equation}
		\#(n,N_\mathbf{0}=n,Y) = \hspace{-3mm}\sum_{X=Y+1}^{\frac{n+Y}{2}} \hspace{-2mm}4\binom{X}{Y}\binom{n-X-1}{X-Y-1} \binom{n-1}{X-1}
	\end{equation}
	The remaining case of $Z=Y=N_\mathbf{0}$, \ie where all movers are mutually separated, gives
	\begin{equation}
		\#(n,N_\mathbf{0},Y=N_\mathbf{0}) = 4\binom{n-1}{N_\mathbf{0}-1}
	\end{equation}
	We can convert $Z$ out of $Y$ zeros, where we get another binomial $\binom{Y}{Z}$ from the number of choices to do so.
	
	The total number of states in the block with $n=\frac{L}{4}$ movers and spins can then be lower-bounded by the sum
	\begin{align}
		\mathcal{D}_{\max}\geq \mathcal{D}_{\max}^- &=\sum^{n-2}_{Y=0}\underbrace{\sum^Y_{Z=0}\binom{Y}{Z}}_{=2^Y}\,\#(n,n,Y)\,.
	\end{align}
	We will approximate the sum over $X$ and $Y$ by integrals over $x$ and $y$, such that $X=xL$ and $Y=yL$. We have $\mathcal{D}^-_{\max}=L^2\int_0^{\frac{1}{4}}dy\int^{\frac{1+2y}{4}}_{y}dx\,I(x,y)$ where we can expand integrand for large $L$ as
	\begin{align}
		I(x,y)&=2^{2+yL}\binom{xL}{yL}\binom{\frac{L}{4}-xL-1}{xL-yL-1} \binom{\frac{L}{4}-1}{xL-1}\\
		&\sim \frac{1}{(\pi L)^{\frac{3}{2}}}\,e^{f_1(x,y)\,L+f_2(x,y)}\,,
	\end{align}
	where the respective functions are
	\begin{align}
		\begin{split}
			f_1(x,y)&=2 (y\!-\!x)\log (x\!-\!y)\!+\!2 x \log \left(y-2 x+\tfrac{1}{4}\right)\\
			&\quad -\!\tfrac{1}{4} \log (4 y\!+\!1\!-\!8 x)\!-\!y \log\tfrac{y (4 y+1-8 x)}{8}
		\end{split}\\
		f_2(x,y)&=-\tfrac{1}{2} \log\tfrac{y (-8 x+4 y+1)}{32}\,.
	\end{align}
	We find the saddle point with $d f_1(x,y)=0$ to be given by $(x_0,y_0)=(\frac{3}{20},\frac{1}{10})$, which is within integration range $0\leq y\leq x\leq \frac{1}{4}+\frac{y}{2}\leq\frac{1}{2}$. We further have $f_1(x_0,y_0)=\tfrac{1}{4}\log{5}$ and $f_2(x_0,y_0)=\log{40}$. Around the saddle, we can approximate the integrand by a Gaussian as
	\begin{align}
		I(x,y)\sim \frac{1}{(\pi L)^{\frac{3}{2}}}\,e^{(f_1(x_0,y_0)-\frac{1}{2}\vec{r}M\vec{r})L+f_2(x_0,y_0)}\,,
	\end{align}
	where $r=(x,y)$ and the matrix $M$ is given by
	\begin{align}
		M_{ij}=-\partial_{r_i}\partial_{r_j}f_1(x_0,y_0)=\begin{pmatrix}
			120 & -80\\
			-80 & 70
		\end{pmatrix}\,.
	\end{align}
	This yields the asymptotics of the integral as
	\begin{align}
		\begin{split}
			\mathcal{D}^{-}_{\max}&\sim\tfrac{L^2}{(\pi L)^{\tfrac{3}{2}}}\sqrt{\tfrac{(2\pi)^2}{L^2\det{M}}}\,e^{f_1(x_0,y_0)\,L+f_2(x_0,y_0)}\sim \tfrac{4}{\sqrt{\pi L}}\,5^{\frac{L}{4}}\,.
		\end{split}
	\end{align}
	
\renewcommand{\thefigure}{S2}		
	\begin{figure}[t!]
		\centering
		\begin{tikzpicture}
			\draw[white] (-4,2.7) rectangle (4,-2.3);
			\foreach \x in {0,...,31}
			{
				\draw (95-10*\x:1.5cm) circle(.7mm);
			}
			\filldraw (95:1.5cm) circle(.7mm);
			\filldraw (95-30:1.5cm) circle(.7mm);
			\filldraw (95-40:1.5cm) circle(.7mm);
			\filldraw (95-70:1.5cm) circle(.7mm);
			\filldraw (95-100:1.5cm) circle(.7mm);
			\filldraw (95-120:1.5cm) circle(.7mm);
			\filldraw (95-130:1.5cm) circle(.7mm);
			\filldraw (95-140:1.5cm) circle(.7mm);
			\filldraw (95-160:1.5cm) circle(.7mm);
			\filldraw (95-170:1.5cm) circle(.7mm);
			\filldraw (95-200:1.5cm) circle(.7mm);
			\filldraw (95-210:1.5cm) circle(.7mm);
			\filldraw (95-230:1.5cm) circle(.7mm);
			\filldraw (95-250:1.5cm) circle(.7mm);
			\filldraw (95-280:1.5cm) circle(.7mm);
			\filldraw (95-290:1.5cm) circle(.7mm);
			\foreach \x in {1,...,6}
			{
				\draw (95-10*\x:1.3cm) node[scale=.7]{$\x$};
			}
			\foreach \x in {2,...,5}
			{
				\pgfmathsetmacro\result{\x*6}
				\draw (95-60*\x:1.25cm) node[scale=.7]{$\pgfmathprintnumber{\result}$};
			}
			\draw (95:1.3cm) node[scale=.7]{$L$};
			\draw (120:1.6cm) node[rotate=30]{$\dots$};
			\draw (90:1.7cm) arc (90:-210:1.7cm);
			\draw (90:1.7cm) -- (90:2.1cm) (30:1.7cm) -- (30:2.1cm) (-30:1.7cm) -- (-30:2.1cm) (-90:1.7cm) -- (-90:2.1cm) (-150:1.7cm) -- (-150:2.1cm) (-210:1.7cm) -- (-210:2.1cm);
			\draw (0,2.1) arc (90:-210:2.1cm);
			\draw[thick,->] (45:2.5cm) arc (45:15:2.5cm);
			\draw[<->] (55:1.75cm) arc (-40:160:1.7mm);
			\begin{scope}[rotate=-60]
				\draw[<->] (55:1.75cm) arc (-40:160:1.7mm);
			\end{scope}
			\begin{scope}[rotate=-120]
				\draw[<->] (55:1.75cm) arc (-40:160:1.7mm);
			\end{scope}
			\begin{scope}[rotate=-180]
				\draw[<->] (55:1.75cm) arc (-40:160:1.7mm);
			\end{scope}
			\begin{scope}[rotate=-240]
				\draw[<->] (55:1.75cm) arc (-40:160:1.7mm);
			\end{scope}
			\draw (45:1.9cm) node[scale=.8]{\checkmark};
			\draw (-15:1.9cm) node[scale=.8]{\checkmark};
			\draw (-75:1.9cm) node[scale=.8]{$\times$};
			\draw (-135:1.9cm) node[scale=.8]{\checkmark};
			\draw (-195:1.9cm) node[scale=.8]{$\times$};
			\draw (3,1.5) node{rotate at};
			\draw (3,1.2) node{each step};
		\end{tikzpicture}
		\caption{We illustrate our Markov simulation where we apply to an initial product state classical ``gates'' of length $6$ that switch the two middle sites with probability $50\%$, provided the rules (2) are satisfied (indicated by above). At each step, we also randomly rotate the position of our gates.}
		\label{fig:markov}
	\end{figure}
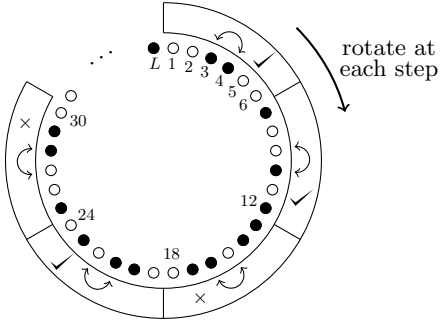

	\subsection{Markov simulation}

 \renewcommand{\thefigure}{S3}	
	\begin{figure}[t]
		\centering
		\begin{tikzpicture}
			\draw (0,0) node{\includegraphics[width=.9\linewidth]{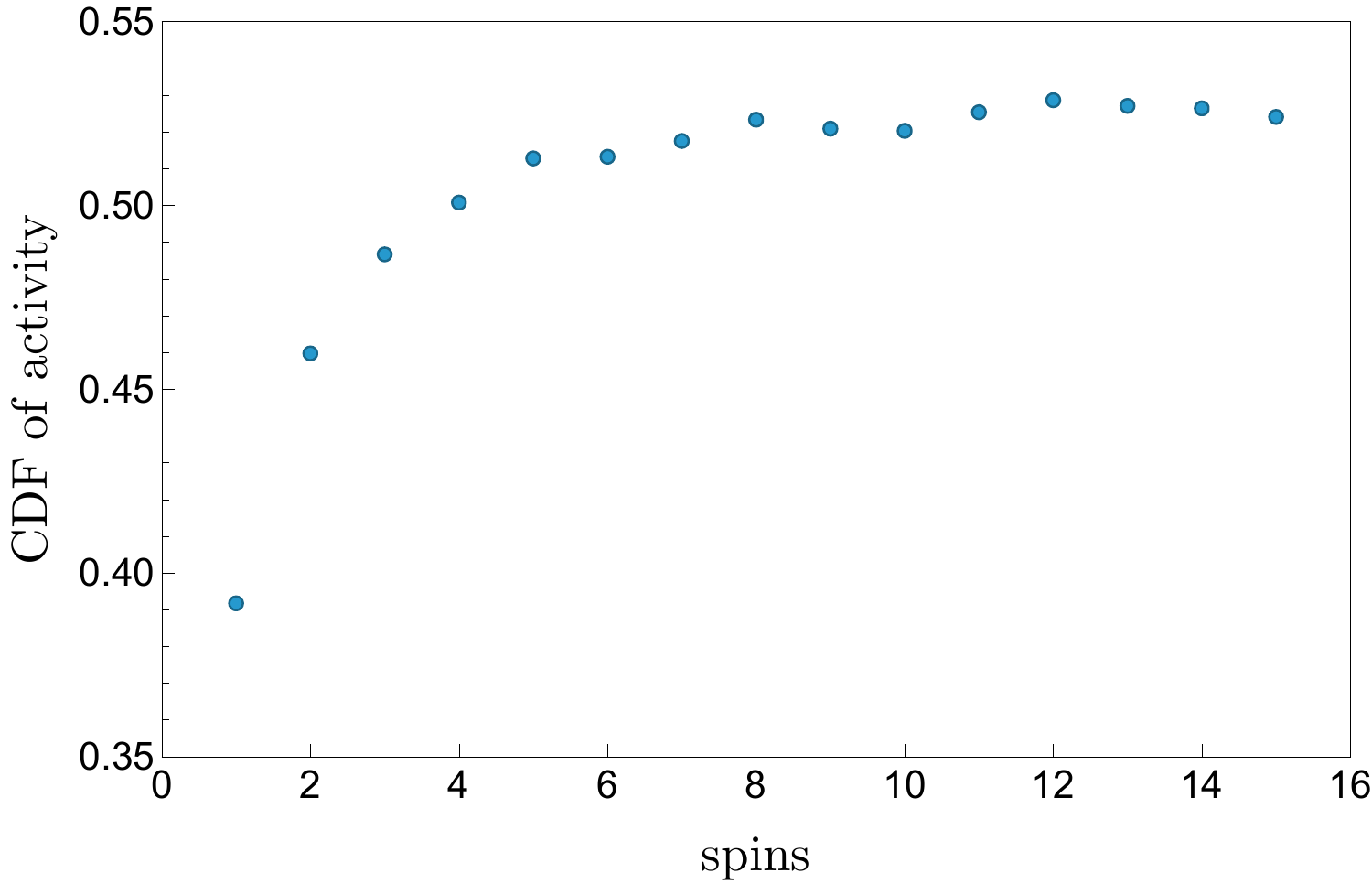}};
			\draw[white] (-4,2.9) rectangle (4,-2.3);
		\end{tikzpicture}
		\caption{Cumulative probability of activity caused by a single mover in a background of spins that a priori cannot generate additional movers for $V_2 \to \infty$. Saturation at a finite value of $\sim 0.53$ indicates that movers are either contained within a finite region or cause the entire system to activate via an avalanche effect. Data points represent the average over $2\cdot10^4$ Markov simulations.}
		\label{fig:CDF V2}
	\end{figure}
	
	Our Markov gates act on four sites in the case of $\hat{H}^{(1)}$ and six sites in the case of $\hat{H}^{(2)}$ by mapping the current set of occupancies to any other allowed state that respects the constrained hopping, including the original one, with equal probability. In our model the only two possibilities are no change in occupancy or a nearest neighbor hopping. First, the system is randomly partitioned into adjacent $k$-site sections ($k=4,6$) in a fashion that is consistent with the given boundary conditions. Then Markov gates are applied to each of these sections. Finally, the resulting strings of occupancies are joined together to form the new state. Figure\,\ref{fig:markov} illustrates the algorithm. Figure\,\ref{fig:CDF V2} shows results for $V_2 \to \infty$ obtained via Markov simulation.
	While Markov simulations of quantum systems do not produce the state that would follow from unitary time evolution, they nevertheless stochastically generate the product basis whose linear combination produces said state. In other word, unitary time evolution would yield states that are contained within the span of states that are generated by a Markov simulation. This is sufficient for our purpose of investigating the block structure of Hamiltonians. In order to ensure that the stochastic nature of Markov simulations does not affect our results, we extend the total number of time steps until past the point where the data becomes independent of the number of steps.

	\subsection{Overlap of active regions for \texorpdfstring{$V_{1,2} \to \infty$}{V12->infty}}
	
	We estimate the overlap between two active regions produced by neighboring movers under the assumption that the statistics of each region is unaffected by the presence of the other one. The mover density $n_0$ only enters into the Poisson distribution of the spacing $\Delta$ between these two movers: $\mathcal{P}^{(n_0)}_\mathrm{sep}(\Delta) = \frac{n_0^{-\Delta} e^{-1/n_0}}{\Delta !}$. Denoting the number of active symbols to the left/right of mover 1 as $n_l/n_r$ and those to the left/right of mover 2 as  $m_l/m_r$ we merely need to sum over all values and distinguish various distinct cases. In the following, $\theta$ denotes the Heaviside step function, $\delta$ the Kronecker delta and $0<\epsilon<1$ can be chosen arbitrarily.
	
	\newpage
	
	\begin{widetext}
		\begin{equation}
			\hspace{-0.2cm}
			\begin{aligned}
				\mathcal{P}(\mathrm{overlap}=x) &= 2 \sum_\Delta \mathcal{P}^{(n_0)}_\mathrm{sep}(\Delta)\sum_{n_l,n_r,m_l,m_r} 2^{-(m_l + m_r + 2)} \cdot 2^{-(n_l + n_r + 2)} \; \theta((m_l + m_r) - (n_l + n_r) + \epsilon) \\
				&\qquad\times \big\{ \theta(m_l + n_r - \Delta - \epsilon) \theta(\Delta + n_l - m_l + \epsilon) \theta(\Delta n_r + \epsilon) \delta_{x,(n_r + m_l - \Delta)}   \\
				&\qquad\qquad+ \theta(\Delta - n_r + \epsilon) \theta(m_l - \Delta - n_l - \epsilon)\delta_{x, (n_l + n_r)} \\
				&\qquad\qquad+ \theta(n_r - \Delta - \epsilon) \theta(n_l + \Delta - m_l + \epsilon) \delta_{x, (m_l + n_r - \Delta)} \\
				&\qquad\qquad+ \theta(n_r - \Delta - \epsilon) \theta(m_l - n_l - \Delta - \epsilon) \theta(m_r - n_r + \Delta + \epsilon) \delta_{x, (n_l + n_r)}  \\
				&\qquad\qquad+ \theta(n_r - \Delta - \epsilon) \theta(m_l - n_l - \Delta - \epsilon) \theta(n_r - \Delta - m_r - \epsilon) \delta_{x, (n_l + \Delta + m_r)} \big\}
			\end{aligned}
		\end{equation}
	\end{widetext}

\end{document}